\def\al{\alpha}
\def\be{\beta}
\def\ga{\gamma}
\def\de{\delta}
\def\ep{\epsilon}

\def\et{\eta}

\def\ka{\kappa}
\def\la{\lambda}

\def\si{\sigma}

\def\ps{\psi}
\def\om{\omega}
\def\Ga{\Gamma}

\def\La{\Lambda}

\def\mn{{\mu\nu}}
\def\cl{{\cal L}}

 \def\expect#1{\langle{#1}\rangle}

 \def\half{{\textstyle{1\over 2}}}

\def\frac#1#2{{\textstyle{{#1}\over {#2}}}} 
\def\lsim{\mathrel{\rlap{\lower4pt\hbox{\hskip1pt$\sim$}} 
\raise1pt\hbox{$<$}}}
\def\gsim{\mathrel{\rlap{\lower4pt\hbox{\hskip1pt$\sim$}} 
\raise1pt\hbox{$>$}}}
\def\sqr#1#2{{\vcenter{\vbox{\hrule height.#2pt 
\hbox{\vrule width.#2pt height#1pt \kern#1pt \vrule width.#2pt}
\hrule height.#2pt}}}}

\def\lrprtnu{\stackrel{\leftrightarrow}{\partial^\nu}} 
\def\lrcov{\stackrel{\leftrightarrow}{D}}

\newcommand{\beq}{\begin{equation}}
\newcommand{\eeq}{\end{equation}}
\newcommand{\bea}{\begin{eqnarray}}
\newcommand{\eea}{\end{eqnarray}}
\newcommand{\rf}[1]{(\ref{#1})}

\documentclass[
    ,final          
  ]
  {aipproc}

\layoutstyle{6x9}

\begin{document}

\title{Theoretical Overview of Lorentz and CPT Violation}

\author{Don Colladay}{
  address={New College of Florida, Sarasota, FL}
}

\begin{abstract}
 In this talk, I discuss some recent theoretical progress concerning
the Lorentz- and CPT-violating extension of the standard model.
The results summarized include the development of an explicit connection between 
noncommutative field theory
and the standard model extension, placement of new bounds in the photon sector, 
calculation of one-loop renormalization beta functions in QED, 
and an analysis of field redefinitions.
\end{abstract}

\maketitle


\section{OVERVIEW}

  For over ten years now there has been active interest in the possibility
that more fundamental theories may induce small violations of Lorentz and
CPT invariance into the standard model at levels accessible
to high precision  experiments \cite{cpt01}.  The original motivation for the idea arose from string theory
\cite{kps} in which higher order field interactions due to the non-local
nature of strings may modify the Lorentz properties of the
vacuum.  The general mechanism developed to model this effect at the level of the standard model
is spontaneous symmetry breaking in which tensor
fields attain a nonvanishing expectation value in the vacuum at low energies.
In fact, the idea of a generic spontaneous symmetry breaking mechanism 
can be applied to generic fundamental theories that reduce to the standard model 
at low energies.  

  A Standard Model Extension (SME) that includes all possible terms arising 
  from a generic spontaneous symmetry breaking mechanism of this type has been constructed 
  \cite{ck}.
These terms may violate Lorentz and/or CPT invariance.
The framework is that of conventional quantum field theory in which standard techniques can
be implemented to calculate the effects of Lorentz and CPT violation on
physical processes.
Sensitive experimental tests of Lorentz and CPT symmetry include accelerator experiments
\cite{hadronexpt,hadronth}, low-energy atomic experiments \cite{pn,eexpt,eexpt2},
and astrophysical tests \cite{cfj, mewes}.

  There is a deep connection between Lorentz invariance and CPT symmetry from the well-known
  CPT theorem as well as the more recent result by Greenberg that CPT violation in fact requires a violation of 
  Lorentz invariance \cite{greenberg}.
In this reference, Greenberg also considers a generic field theory in which one tries to introduce
a separate mass for the particle and antiparticle states.
He shows that there is necessarily a violation of locality as well as a violation of 
coordinate Lorentz invariance in such theories.
This implies that different observers would not be able to make consistent calculations
in such a theory.
  Therefore, bounds of CPT symmetry can be interpreted as bounds on Lorentz invariance.
Note that Lorentz violation does not necessarily imply CPT violation as can be seen from
explicit terms of this type in the SME.
  
  To begin, I will discuss the construction of the 
SME as well as some motivation for its development.  
Following this introduction, I will give a summary of four theoretical papers 
that have significantly developed the framework involved.
Other papers in this proceedings include an analysis of a supersymmetric
generalization \cite{mberger} and Lorentz violation induced time variation
of physical constants \cite{rlehnert}.  
  
\section{INTRODUCTION TO LORENTZ AND CPT VIOLATION}

As mentioned in the previous section, there is good theoretical motivation 
for the possibility that Lorentz and CPT invariance may be 
an approximation at low-energy scales.  
In addition to the theoretical motivation, 
many experiments that involve high precision tests of 
relativity require a common framework within which to compare bounds on various 
types of physical measurements.  
Having an explicit theory in terms of the fundamental fields of the standard
model allows different experiments to compare bounds on parameters and get a 
quantitative handle on effects that they are sensitive to.
For example, the photon spectropolarimetry measurements place a bound on Lorentz 
violation in the photon sector that is sixteen orders of magnitude more stringent
than the Gamma ray burst and pulsar data \cite{mewes}.  
This will be discussed later in the talk.

The mechanism used to generate the SME is 
spontaneous symmetry breaking applied to fields with tensor indices.
This mechanism is analogous to the Higgs mechanism in which a scalar
field gains a vacuum expectation value and generates masses for the 
standard model particles.
In the case of a tensor field ($B^\mu (x)$ for example) containing 
Lorentz indices, a nonzero expectation value will select out a specific
direction in spacetime breaking Lorentz invariance spontaneously.
Potentials for tensor fields are absent in conventional renormalizable
field theories but can occur in low-energy field expansions of more 
fundamental theories such as string field theory \cite{kps}.
Couplings between these tensor fields and standard model
particles (such as $B_\mu \overline \psi \ga^5 \ga^\mu \psi$) induce
violations of Lorentz invariance in the low-energy effective
theory due to the spontaneous symmetry breaking (for example $\expect{B_\mu} \ne 0$).

The SME {\cite {ck}} consists of all such terms
arising from couplings between standard model fields and background tensor fields. 
In general, there are terms in the SME that are nonrenormalizable
and terms that violate gauge symmetries.
To simplify, it is useful to restrict this very general theory of Lorentz and CPT 
violation to satisfy SU(3) $\times$ SU(2) $\times$ U(1) gauge invariance  and power counting
renormalizability.
Restricting further to spacetime independent expectation values generates 
the minimal SME that is useful for quantifying leading order
corrections to experiments.

As an example, the QED sector of the minimal SME
is given here.  
The QED extension is obtained by
restricting the minimal SME to the electron and photon
sectors. 
The electron terms are\footnote{Additional correction terms consistent with $U(1)$
symmetry of the electromagnetic sector, but not with SU(2) symmetry of the full
electroweak sector are often also included.  With this relaxed condition, the terms $e^\nu + i f^\nu \ga^5 + 
\frac 1 2 g^{\la\mu\nu}\si_{\la\mu}$ may be added into the definition of $\Ga^\nu$. }
\begin{equation}
\cl_e =
\half i \overline{\ps}  \Ga^\nu \lrcov_\nu \ps 
- \overline{\ps}  M  \ps
\quad ,
\label{eleclag}
\end{equation}
where $\Ga$ and $M$ denote
\beq
{\Ga}^{\nu}  ={\ga}^{\nu}+  c^{\mu \nu} 
{\ga}_{\mu}+  d^{\mu \nu} {\ga}_{5} {\ga}_{\mu}
\quad ,
\eeq
\beq
 M  = m+  a_{\mu}  {\ga}^{\mu}+  b_{\mu}
 {\ga}_{5} {\ga}^{\mu}+\frac{1}{2} H^{\mu \nu}
 {\sigma}_{\mu \nu}
\quad .
\eeq 
The parameters $a$, $b$, $c$, $d$, and $H$ are fixed background
expectation values of tensor fields
that break conventional particle Lorentz symmetry.
The photon terms are
\beq
\cl_\ga  = - {1 \over 4} F^\mn F_\mn  
- {1 \over 4} (k_F)_{\ka\la\mu\nu} F^{\ka\la} F^{\mn}
+ {1 \over 2} (k_{AF})^\ka \ep_{\ka\la\mn} A^\la F^\mn
\quad ,
\label{photlag}
\eeq
where $k_F$, $k_{AF}$ are the fixed background tensor fields.

Next, I will discuss several new theoretical results that have emerged
over the past two years that are associated with the general SME.

\section{CONNECTION TO NONCOMMUTATIVE FIELD THEORY}

There has been much interest recently in the possibility that the coordinates
used to parameterize the standard model fields may not commute.  Such a 
situation can arise naturally in the low energy limit of certain string 
theories \cite{stringnoncom}.  In this case, the nonvanishing commutators
can take the special form 
\beq
[x^\mu, x^\nu] = i \theta^{\mn}
\eeq
where the parameters $\theta$ violate Lorentz invariance as they are
fixed background parameters.
It has been shown \cite{lane} that {\it any} realistic theory of noncommutative
geometry must be physically equivalent to a subset of the SME.
The proof relies on the existence of a correspondence between the fields on noncommutative
coordinates and conventional fields on commutative coordinates called the Seiberg-Witten 
map \cite{swmap}.
The result follows by using this map to identify the appropriate Lorentz-violating 
extension parameters that are present in the resulting theory.
More recently, the map has been applied to the entire standard model \cite{wess}.
The authors find terms that are consistent with a subset of the SME as expected.
As an explicit example, a noncommutative version of QED developed in \cite{lane} 
is discussed here.

One way of implementing the noncommutative structure of the 
underlying coordinates is to promote an established theory to a noncommutative
one using the Moyal $\star$ product representation 
\beq
(f \star g)(x) = \exp{(\frac 1 2 i \theta^\mn \partial_{x^\mu} \partial_{y^\nu})f(x) g(y)|_{x=y}}
\quad ,
\eeq
for multiplication of the fields.
Noncommutative QED can then be constructed using this multiplication as
\beq
\cl = \frac i 2 \overline{\hat \psi} \star \ga^\mu \stackrel{\leftrightarrow}{\hat D_\mu} \hat \psi
- m \overline{\hat \psi} \star \hat \psi - {\frac 1 {4 q^2}}\hat F_{\mn} \star \hat F^\mn
\quad .
\eeq
These noncommutative fields $(\hat \psi, \hat A)$ satisfy unconventional gauge transformations and 
do not correspond to the conventional electrons and photons as described in
the framework of conventional quantum field theory.  
Application of the Seiberg-Witten map \cite{swmap} (to lowest order in $\theta$) 
\beq
\hat A_\mu = A_\mu - \frac 1 2 \theta^{\al \be}A_\al (\partial_\be A_\mu + F_{\be \mu})
\quad ,
\eeq

\beq
\hat \psi = \psi - \frac 1 2 \theta^{\al \be}A_\al \partial_\be \psi
\quad ,
\eeq
must be used to identify the relevant 
corrections to the standard electrodynamic fields $(\psi, A)$ in a form that can be directly 
compared to experimental results.
The resulting effective QED theory becomes (to first order in $\theta$)
\bea
\cl = & & \cl_0  -  \frac i 8 q \theta^{\al\be} F_{\al\be} \overline \psi  \ga^\mu
\stackrel{\leftrightarrow}{D_\mu} \ps
+ \frac 1 4 iq \theta^{\al\be} F_{\al\mu}
\overline \psi \ga^\mu \stackrel{\leftrightarrow}{D_\be} \psi \nonumber \\
& & + \frac 1 4 m q \theta^{\al\be} F_{\al\be} \overline \psi \psi + (F^3 \cdots)
\quad .
\eea
The correction terms correspond to nonrenormalizable terms in the 
SME\footnote{The possibility of renormalizable terms
emerging from loop corrections has been explored \cite{anis}.}.  
It is possible to examine experiments that occur in constant background
electromagnetic fields using $F^\mn \rightarrow f^\mn + F^\mn$ where $f^\mn$
is constant.
With this substitution, a specific subset of the terms in the minimal SME are recovered.
These are 
\beq
\cl = \cl_0 + \frac i 2 c_\mn 
\overline \psi \ga^\mu \stackrel{\leftrightarrow}{D^\nu}\psi
- \frac 1 4 (k_F)_{\al \be\ga\de} F^{\al\be}F^{\ga\de}
\quad ,
\eeq
with
\beq
 c_\mn = - \frac 1 2 q f_\mu^\la \theta_{\la\nu} ~~ ; ~~~ 
(k_F)_{\al\be\ga\de}  = - q f_\al^\la \theta_{\la\ga}\et_{\be\de} + \cdots
\quad .
\eeq
Atomic experiments in constant $B$ fields can then be used to bound the 
noncommutative parameters at the level
\beq
|\theta^{ij}|<(10{\rm ~TeV})^{-2}
\quad .
\eeq
Effects of noncommutative geometry on photon propagation in constant 
background fields have also been considered \cite{gural}.

\section{ONE LOOP RENORMALIZABILITY OF QED SECTOR}

The next result concerns the explicit analysis of the 
one-loop renormalizability of the
QED sector of the minimal SME \cite{pick}.
Results included in this reference include:
\begin{itemize}
\item{Generalized Furry theorem is established showing that the three and four point 
photon vertices generate a finite contribution to one-loop Green's functions.}
\item{Multiplicative renormalization holds at one loop provided
the Lorentz-violating constants are properly renormalized.}
\item{The beta functions have been calculated and the renormalization group
was used to examine the running of the violation parameters.}
\end{itemize}
The modified vertices and propagators can be extracted from the lagrangian
for extended QED given in equations \rf{eleclag} and \rf{photlag}.
For example, the electron-photon vertex will contribute a factor of
$-i q \Ga^\mu = -i q(\ga^\mu + \ep^\mu)$ where $\ep^\mu$ is a small 
perturbative correction that depends on the Lorentz and CPT violating terms.

The running of the coupling constants were calculated using renormalization
group techniques.  
They are found to depend on various anomalous powers
of the parameter
\beq
Q(\mu) = 1 - {q_0^2 \over 6 \pi^2} \ln{\mu \over \mu_0}
\quad ,
\eeq
that controls the usual running of the QED charge according to 
$q = Q^{-1} q_0$.
In the above expressions, $\mu$ is the renormalized mass scale while $\mu_0$
is a reference scale at which the boundary conditions on the parameters are applied.
As an example, 
the $a$ parameters run according to
\beq
a^\mu = a_0^\mu - m_0 (1 - Q^{9/4})e_0^\mu
\quad ,
\eeq
while the $c$ parameters run as
\beq
c^{\mn} = c_0^{\mn} - \frac 1 3 (1 - Q^{-3})(c_0^\mn + c_0^{\nu\mu} - (k_{F})_{0~~~~\al}^{~\mu\nu\al})
\quad .
\eeq
If the parameters are assumed to be unified at the Planck scale, a naive running 
to low energies indicates that the parameters can differ by 2 to 3 orders of magnitude
at the low-energy scale.  This result emphasizes the necessity of independently measuring all of the parameters
that control Lorentz violation as they may be very different in size.

\section{NEW BOUNDS ON PARAMETERS IN THE PHOTON SECTOR}

Various cosmological experiments have already placed stringent bounds on
the CPT violating photon terms \cite{cfj}.
In addition, there are theoretical reasons to suggest that these terms are
exactly zero \cite{ck, kle}.
However, there is no such theoretical bias concerning 
the CPT even photon terms and they have recently been analyzed in \cite{mewes}.
In this reference, an explicit analogy is constructed between photon propagation in a classical
anisotropic medium and photon propagation in a Lorentz-violating background
field.
The formalism provides a clean way of extracting bounds on all of the CPT-even
parameters using both astrophysical
and lab based photon propagation experiments.

The relevant term $k_F$ modifies the Maxwell equations according to
\beq
\partial_\al F_{\mu}^{\al} = -  (k_F)_{\mu\al\be\ga} 
\partial^\al F^{\be \ga}
\quad , \quad \partial_\mu \tilde F^\mn = 0
\eeq
Note that the homogeneous equations are unmodified as these only 
depend on the definition of $F^\mn = \partial^\mu A^\nu - \partial^\nu A^\mu$.
To construct the analogy with anisotropic media, fields $\vec D$ and $\vec H$
are defined according to
\beq
\left(
\begin{array}{c}
  \vec D \\
  \vec H
\end{array}  
\right)
=
\left(
\begin{array}{cc}
  1 + \ka_{DE} &  \ka_{DB} \\
  \ka_{HE} &  1 + \ka_{HB}
\end{array}
\right)
\left(
\begin{array}{c}
  \vec E \\
  \vec B
\end{array}  
\right)
\quad ,
\eeq
where the various $\ka$ quantities are $2\times 2$ constant matrices
depending on the $k_F$ parameters.
Using this definition, the modified Maxwell equations take the 
conventional form of
\bea
\vec \nabla \times \vec H - \partial_0 \vec D = 0 ~& , & 
\quad \vec \nabla \cdot \vec D = 0 \nonumber \\
\vec \nabla \times \vec E + \partial_0 \vec B = 0 ~& , &
\quad \vec \nabla \cdot \vec B = 0 \nonumber
\quad .
\eea
The form of these equations implies that standard techniques can be used to 
solve the equations of motion.

The violation terms can be divided into ones that cause birefringence and ones
that do not.  
Birefringence measurements can be performed with high sensitivity using astrophysical 
measurements, while the other terms can be bounded using various resonant cavity 
experiments.  I will focus here on two types of astrophysical tests analyzed in \cite{mewes}. 
The first involves gamma ray bursts and pulsars while the second involves
spectropolarimetry measurements.  Refer to \cite{mewes} for details concerning the 
cavity experiments.

Gamma ray bursts and pulsars produce narrow pulses of radiation that propagate 
large distances.
Birefringence implies a velocity difference between the eigenmodes of propagation
yielding a spreading of the pulse width in time of $\Delta t \approx \Delta v L$,
where $L$ is the distance to the source.
Using fifteen different sources, a conservative bound of $|k_F|<3 \times 10^{-16}$
has been placed on CPT-even parameters that cause birefringence.  

Much more accurate bounds have been placed on the same parameters using 
spectropolarimetry data.  
It is difficult to determine the polarization of most astrophysical sources accurately,
so a technique of searching for a specific wavelength dependence in the polarization
rotation was implemented.
A detailed analysis of the modified Maxwell equations shows that
the polarization shift due to birefringence is proportional to the inverse of the wavelength.
Combining this fact with
the extremely precise time resolution of phase shift time scales yields a bound of
$|k_F|<2 \times 10^{-32}$ on the same parameters that are bounded using pulse
broadening analysis.
This points out the necessity of having a specific theory to calculate
explicit bounds on Lorentz symmetry.  
Spectropolarimetry bounds are far more stringent and they require a detailed 
knowledge of the form of the modified Maxwell equations as is given in the
SME. 
A simple phenomenological correction to the dispersion relation 
is not sufficient for a comparable analysis.

\section{FIELD REDEFINITIONS AND LORENTZ VIOLATION}

As the final development discussed in this talk, I will present work done regarding
the physical nature of various terms present in the SME \cite{cmcd}.
Some terms that are included in the lagrangian \rf{eleclag} can be eliminated
using suitable redefinitions of the spinor components.  
Other terms can be moved to different sectors of the theory.
In general, one can define a set of equivalence classes for lagrangians in the
SME by relating elements that are connected by redefinitions
of the field components.
It is not necessary for the redefinitions to be covariant, so the equivalence
class of Lagrangians associated to the standard one contains many terms 
that apparently violate Lorentz or CPT invariance.

To illustrate the general procedure, we start with the conventional lagrangian
for QED 
\beq
\cl[\psi] = \frac i 2 \overline \psi \gamma^\mu \stackrel{\leftrightarrow}{D_\mu}
\psi  -m \overline \psi \psi 
\quad ,
\eeq
and apply a redefinition of the spinor field of the form
\beq
\psi(x) = [1 + f(x,\partial)] \chi(x) \quad ,
\eeq
generating a new lagrangian $\cl[\chi]$ that may apparently violate Lorentz invariance.
As an example, let $f = \frac 1 2 v_\mu \ga^\mu$ where $v_\mu$ are real constants.
To lowest order in $v$, the lagrangian expressed in terms of $\chi$ is
\beq
\cl[\chi] = \cl_0 + \frac i 2 v_\mu 
\overline \chi \stackrel{\leftrightarrow}{D^\mu} \chi
+ m v_\mu \overline \chi \ga^\mu \chi
\quad .
\label{fakelv}
\eeq
If one naively assumes the standard action of SL(2,C)  on the spinors of 
\beq
\chi^\prime(x^\prime) = S(\La)\chi(x) = e^{\frac i 4 \om_\mn \si^\mn} \chi(x)
\eeq
then $\cl$ is not covariant.
However, it is in fact covariant under the modified action of 
\beq
\tilde S(\Lambda) = 
e^{-\frac 1 2 v_\mu \ga^\mu}
S(\Lambda)e^{\frac 1 2  v_\mu \ga^\mu}
\quad ,
\eeq
which is related to the standard action by a similarity transformation. 
This logic can be applied in reverse to conclude that any lagrangian of the form
\rf{fakelv} does not in fact violate Lorentz invariance because the fields
can be appropriately redefined\footnote{Note that an interaction term between 
a fermion with a free lagrangian of this form and another particle with
conventional transformation properties may not be invariant under the
redefinition in which case the parameter $v_\mu$ would be physical.}.

Other redefinitions can involve derivatives and are more complicated.
For example, letting $f = C_\mn x^\mu \partial^\nu$ yields a transformed
lagrangian of (lowest order in $C$)
\beq
\cl[\chi] = \cl_0 +  C_\mn x^\mu
\partial^\nu
\cl_0 + \frac i 2 C_\mn 
\overline \chi \gamma^\mu \lrprtnu \chi
\quad .
\eeq
The second term in this expression is a total derivative
up to the term $C^\mu_{~\mu} \cl_0$, a term that simply scales the lagrangian.
The third term in the expression is the form of the $c$ corrections to
$\Ga$ in equation \rf{eleclag}.
This transformation is equivalent to a change of coordinates according to
\beq
\psi(x) = (1 + C_\mn x^\mu \partial^\nu)\chi(x) \approx 
\chi(x + C \cdot x) = \chi(x^\prime)
\quad ,
\eeq
where the new coordinates have a metric of 
$g^\mn = \eta^\mn + C^\mn + C^{\nu\mu}$.
The antisymmetric piece does not alter the metric and
corresponds to a conventional Lorentz transformation.
The alteration in the form of the lagrangian in this case is
compensated by the appropriate element of SL(2,C) for the transformation.
The symmetric piece is more interesting as it skews the coordinate 
system.  
This can be compensated for using the vierbein formalism of general
relativity, but a redefinition of the metric in the photon sector will also be
required.
Therefore these terms may be eliminated from the electron sector, but
they will reappear as corrections in the photon sector.
One can understand this result physically as the necessity of
using the propagation properties of some particular field to define
the coordinate system basis.  
Once this system is chosen, it is then necessary to measure the propagation
properties of other particles with respect to it.  
Any incompatibility in the interactions will lead to a 
potentially observable violation
of Lorentz invariance in the overall theory.

Terms that can be altered by redefinitions can be reexpressed as appropriate
linear combinations such that the terms that are invariant under the field
redefintions will correspond to the 
physically observable parameters.
This can be used to significantly simplify models containing Lorentz
and CPT violation by reducing the number of parameters that must be
included in the calculations.
A more complete analysis is performed in \cite{cmcd}.

\section{SUMMARY}

In this talk, an overview of recent theoretical progress pertaining
to the theory of Lorentz and CPT violation has been presented.

An explicit connection has been made between a subset of the 
SME and physically realistic
theories involving noncommutative field theory.  
In fact, any theory that violates Lorentz or CPT invariance must reduce to
a subset of the general SME provided that
corrections to conventional standard model fields are considered and
different observers can make consistent calculations 
regarding physical processes.
The extension therefore provides a very robust framework within which
violations of Lorentz and CPT symmetry can be analyzed.

One loop renormalizability in the minimal QED extension has been
explicitly established.
The beta functions indicate a variety of runnings for the various
Lorentz- and CPT-violating coupling constants.
As a result of the renormalization group analysis, it is possible
that parameters that are unified at the Planck scale can differ
by a few orders of magnitude at the low-energy scale.

Some apparent violations in the SME can 
be removed by appropriate field redefinitions.
In addition, some parameters can be moved to different sectors
using other types of redefinitions.
The terms that cannot be altered by redefinitions therefore
provide the physically measurable quantities, in accordance with
the explicit calculations performed for experimental observables.





\begin{thebibliography}{99}

\bibitem{cpt01}
For a summary of recent theoretical work and
experimental tests
see, for example,
{\it CPT and Lorentz Symmetry}, V.A.\ Kosteleck\'y, ed., 
World Scientific, Singapore, 1999; 
{\it CPT and Lorentz Symmetry II}, V.A.\ Kosteleck\'y, ed.,
World Scientific, Singapore, 2002.

\bibitem{kps}
V.A.\ Kosteleck\'y and S.\ Samuel,
Phys.\ Rev.\ D {\bf 39}, 683 (1989);
{\it ibid.} 
{\bf 40}, 1886 (1989);
Phys.\ Rev.\ Lett.\ {\bf 63}, 224 (1989);
{\it ibid.} 
{\bf 66}, 1811 (1991);
V.A.\ Kosteleck\'y and R.\ Potting,
Nucl.\ Phys.\ B {\bf 359}, 545 (1991);
Phys.\ Lett.\ B {\bf 381}, 89 (1996);
Phys.\ Rev.\ D {\bf 63}, 046007 (2001); 
V.A.\ Kosteleck\'y, M.\ Perry, and R.\ Potting,
Phys.\ Rev.\ Lett.\ {\bf 84}, 4541 (2000). 

\bibitem{ck} 
D.\ Colladay and V.A.\ Kosteleck\'y,
Phys.\ Rev.\ D {\bf 55}, 6760 (1997);
Phys.\ Rev.\ D {\bf 58}, 116002 (1998);
V.A.\ Kosteleck\'y and R.\ Lehnert,
Phys.\ Rev.\ D {\bf 63}, 065008 (2001).


\bibitem{hadronexpt}
KTeV Collaboration,
H.\ Nguyen, in Ref.\ \cite{cpt01};
OPAL Collaboration, 
R.\ Ackerstaff 
{\it et al.},
Z.\ Phys.\ C {\bf 76}, 401 (1997);
DELPHI Collaboration,
M.\ Feindt 
{\it et al.},
preprint DELPHI 97-98 CONF 80 (1997);
BELLE Collaboration,
K.\ Abe
{\it et al.},
Phys.\ Rev.\ Lett.\ {\bf 86}, 3228 (2001);
FOCUS Collaboration,
J.M.\ Link {\it et al.}, hep-ex/0208034.

\bibitem{hadronth}
D.\ Colladay and V.A.\ Kosteleck\'y,
Phys.\ Lett.\ B {\bf 344}, 259 (1995);
Phys.\ Rev.\ D {\bf 52}, 6224 (1995);
Phys.\ Lett.\ B {\bf 511}, 209 (2001);
V.A.\ Kosteleck\'y and R.\ Van Kooten,
Phys.\ Rev.\ D {\bf 54}, 5585 (1996);
V.A.\ Kosteleck\'y,
Phys.\ Rev.\ Lett.\ {\bf 80}, 1818 (1998);
Phys.\ Rev.\ D {\bf 61}, 016002 (2000);
{\bf 64}, 076001 (2001);
N.\ Isgur {\it et al.}
Phys.\ Lett.\ B {\bf 515}, 333 (2001).

\bibitem{pn}
L.R.\ Hunter
{\it et al.},
in V.A.\ Kosteleck\'y, ed.,
\it CPT and Lorentz Symmetry, \rm
World Scientific, Singapore, 1999;
D.\ Bear
{\it et al.},
Phys.\ Rev.\ Lett.\ {\bf 85}, 5038 (2000);
D.F.\ Phillips
{\it et al.},
Phys.\ Rev.\ D {\bf 63}, 111101 (2001);
M.A.\ Humphrey 
{\it et al.},
Phys.\ Rev.\ A {\bf 62}, 063405 (2000);
V.A.\ Kosteleck\'y and C.D.\ Lane,
Phys.\ Rev.\ D {\bf 60}, 116010 (1999);
J.\ Math.\ Phys.\ {\bf 40}, 6245 (1999);
R.\ Bluhm {\it et al.},
Phys.\ Rev.\ Lett.\ {\bf 88}, 090801 (2002).

\bibitem{eexpt}
H.\ Dehmelt 
{\it et al.},
Phys.\ Rev.\ Lett.\ {\bf 83}, 4694 (1999);
R.\ Mittleman 
{\it et al.},
Phys.\ Rev.\ Lett.\ {\bf 83}, 2116 (1999);
G.\ Gabrielse 
{\it et al.},
Phys.\ Rev.\ Lett.\ {\bf 82}, 3198 (1999);
R.\ Bluhm {\it et al.},
Phys.\ Rev.\ Lett.\ {\bf 82}, 2254 (1999);
Phys.\ Rev.\ Lett.\ {\bf 79}, 1432 (1997);
Phys.\ Rev.\ D {\bf 57}, 3932 (1998).

\bibitem{eexpt2}
B.\ Heckel
in Ref.\ \cite{cpt01}; 
R.\ Bluhm and V.A.\ Kosteleck\'y,
Phys.\ Rev.\ Lett.\ {\bf 84}, 1381 (2000).

\bibitem{cfj}
S.M. Carroll, G.B. Field, and R. Jackiw, 
Phys.\ Rev.\ D {\bf 41}, 1231 (1990).

\bibitem{mewes}
V.A. Kosteleck\'y and M. Mewes,
Phys.\ Rev.\ Lett.\ {\bf 87} 251304(2001);
Phys.\ Rev.\ D {\bf 66} 056005 (2002).

\bibitem{greenberg}
O.W.\ Greenberg,
Phys.\ Rev.\ Lett.\ {\bf 89}, 2316021 (2002). 

\bibitem{mberger}
M. Berger, this proceedings (hep-ph/0212353).

\bibitem{rlehnert}
R. Lehnert, this proceedings;
V. A. Kosteleck\'y, R. Lehnert, M. Perry, (astro-ph/0212003).

\bibitem{stringnoncom}
A. Connes, M. Douglas, and A. Schwartz, JHEP {\bf 02} 003 (1998);
For recent reviews see N.A. Nekrasov, hep-th/0011095;
A. Konechny and A. Schwartz, hep-th/0012145; J. A. Harvey, hep-th/0102076.

\bibitem{lane}
S. Carroll {\it et al.}, 
Phys.\ Rev.\ Lett.\ {\bf 87}, 141601 (2001). 

\bibitem{swmap}
N. Seiberg and E. Witten,
JHEP {\bf 09} 032 (1999);
A.A. Bichl {\it et al.}, hep-th/0102103.

\bibitem{wess}
X. Calmet {\it et al.}, 
Eur.\ Phys.\ J.\ {\bf C23} 363 (2002).

\bibitem{anis}
A.\ Anisimov {\it et al.}, hep-ph/0106356.


\bibitem{gural}
Z.\ Guralnik {\it et al.},
Phys.\ Lett.\ B {\bf 517}, 450 (2001);

\bibitem{pick}
V.A. Kosteleck\'y, C. Lane, and A. Pickering,
Phys.\ Rev.\ D {\bf 65} 056006 (2002).

\bibitem{kle} 
V.A.\ Kosteleck\'y and R.\ Lehnert,
Phys.\ Rev.\ D {\bf 63}, 065008 (2001).

\bibitem{cmcd}
D. Colladay and P. McDonald, 
J.\ Math.\ Phys.\ {\bf 43} 3554 (2002).

\end{thebibliography}


\end{document}